\documentclass[12pt,reqno]{article}
\usepackage{amsfonts}
\usepackage{amsmath}
\usepackage{amsbsy}
\usepackage{graphicx}
\usepackage{amssymb,latexsym}
\numberwithin{equation}{section}
\usepackage{cite}

\begin{document}

\title{Matter coupling in Minimal Massive 3D Gravity and spinor-matter interactions in exterior algebra formalism}
\begin{titlepage}
\author{ Hakan Cebeci\footnote{E.mail:
hcebeci@eskisehir.edu.tr ; hcebeci@gmail.com} , \\
{\small Department of Physics, Eski\c{s}ehir Technical University, 26470 Eski\c{s}ehir, Turkey} }

\date{ }

\maketitle

\bigskip

\begin{abstract}
\noindent In this work, by employing the exterior algebra formalism, we study the matter coupling in Minimal Massive 3D Gravity (MMG) by first considering that the matter Lagrangian is connection-independent and then considering that the matter coupling is connection-dependent. The matter coupling in MMG has been previously investigated in the work Arvanitakis {\it et al} (2014 {\it Class. Quantum Grav.} {\bf 31} 235012) in tensorial notation where the matter Lagrangian is considered to be connection-independent. In the first part of the present paper, we revisit the connection-independent matter coupling by using the language of differential forms. We derive the MMG field equation and construct the related source 2-form. We also obtain the consistency relation within this formalism. Next, we examine the case where the matter Lagrangian is connection-dependent. In particular, we concentrate on the spinor-matter coupling and obtain the MMG field equation by explicitly constructing the source term. We also get the consistency relation that the source term should satisfy in order that spinor-matter coupled MMG equation be consistent.

\vspace{1cm}

\end{abstract}

\end{titlepage}

\section{Introduction}

Einstein theory of gravity with a cosmological constant in $ (2+1) $ dimensions has no propagating degrees of freedom such that the theory in its own is non-dynamical. Hence, the massive gravity models have attracted much attention over the years. The first one of these models is the Topologically Massive Gravity (TMG) which has been constructed by adding gravitational Chern-Simons term to Einstein theory of gravity with a negative cosmological constant \cite{deser_1,deser_2}. When a linearization of field equations has been performed about $ AdS_{3} $ vacuum, it has been seen that TMG has one massive propagating graviton mode. This has led to a research about whether there exists a holographically dual conformal field theory (CFT) on 2-dimensional boundary by making TMG a possible candidate of a consistent quantum theory of gravity. However, the use of the methods outlined in \cite{brown} has shown that TMG has undesirable features in the sense that if the bulk graviton mode has positive energy, one of the central charges calculated by Virasoro algebra is negative leading to a non-unitary CFT on the boundary \cite{kraus}. This problem is identified as  bulk versus boundary clash. To circumvent the problem, a new massive gravity theory, dubbed New Massive Gravity (NMG) \cite{bergshoeff_1,bergshoeff_2}, has been proposed where the new theory involves parity-even higher order curvature terms contrary to TMG that has parity-odd gravitational Chern-Simons term. Yet, the bulk-boundary clash problem has not been resolved satisfactorily. Recently, to overcome the bulk-boundary clash problem, Minimal Massive Gravity theory has been introduced \cite{bergshoeff_3}. This new theory has same gravitational degrees of freedom as TMG such that the theory involves a single propagating bulk graviton as well. It has been shown that MMG solves bulk-boundary clash problem within a certain range of its parameters. It has been seen that bulk graviton has positive energy while there exist positive central charges for dual CFT such that the unitarity is not lost. In fact, an elaborate investigation of unitarity structure of MMG has proven that the theory is unitary within a certain range of its parameters \cite{bergshoeff_3,arvanitakis_1}. Having shown that the MMG theory is unitary and it solves bulk-boundary clash problem, some higher curvature extensions of MMG have been investigated by studying linearized field equations of the extended theory around $ AdS_{3} $ spacetime as well \cite{setare,altas_1}. To explore further physical properties of the theory, canonical structure of MMG has been constructed while introducing its canonical Hamiltonian and calculating energy and angular momentum of BTZ solutions within this formalism \cite{yekta}. On the other hand, in a very recent study \cite{deger_1}, it has been shown that the existence of single massive gravitating mode in both TMG and MMG is related to spontaneous symmetry breaking in Hietarinta-Maxwell Chern-Simons gravity theory constructed via Hietarinta-Maxwell algebra.  

In recent years, it has been demonstrated that MMG possesses the similar type of solutions as TMG. Being a physically viable massive gravity theory, it has been proven that MMG has a unique maximally-symmetric vacuum for a particular value of cosmological parameter referring to a solution at the merger point. In addition, it has been shown that the theory accepts Kaluza-Klein $ AdS_{2} \times S^{1} $ type black hole solutions at the merger point as well as warped $ AdS_{3} $ vacuum solutions \cite{arvanitakis_2}. Then in \cite{alishahiha}, wave and warped AdS solutions of vacuum MMG has been investigated by proving that the theory admits logarithmic solutions at the critical points as well. Later, conformally-flat vacuum solutions of MMG have been introduced where Cotton tensor vanishes for such spacetimes \cite{arvanitakis_3}. Also in recent years, according to Segre-Petrov classification, type $ D $ and type $ N $ solutions with constant curvature scalar have been worked out \cite{altas_2}. Furthermore, Kundt solutions have been obtained in \cite{deger_2} possessing constant scalar invariant spacetimes which correspond to deformations of round and warped AdS, as well as non-constant scalar invariant solutions at the merger point. Very recently, in \cite{deger_3}, different classes of homogeneous solutions including stationary Lifshitz solution with Lifshitz parameter $ z = - 1 $ have been discovered. In addition, in \cite{sarioglu}, static and stationary circularly symmetric solutions have been constructed at the merger point.

Apart from discovering vacuum physical properties of MMG and the solutions that theory admits, matter coupling in MMG also deserves for its further study. The matter coupling in MMG theory has been first investigated in \cite{arvanitakis_2} by using tensorial notation where in that work the authors consider that the matter Lagrangian is assumed to depend on the metric together with some matter fields but not on the connection. It has been found that matter coupling in MMG differs from matter coupling in TMG where for the Iatter it is seen that matter-coupled TMG equation includes only matter stress-energy tensor as a source. On the other hand, as is also explicitly illustrated in \cite{arvanitakis_2} within the use of tensorial notation, the matter coupling in MMG has interesting consequences such that the (on-shell) consistency of matter-coupled MMG field equation requires a source tensor that involves covariant derivatives of stress-energy tensor in addition to terms that are quadratic in stress-energy tensor. With a completely different source term compared to TMG, the matter-coupled MMG field equation admits some exact solutions for certain matter Lagrangians. For instance in \cite{arvanitakis_2}, Friedmann-Lemaitre-Robertson-Walker (FLRW) type cosmological solutions have been obtained for an ideal fluid. Later in \cite{arvanitakis_3}, the black string and domain wall solutions have been presented for the scalar matter Lagrangian involving a potential term. In addition in a recent work \cite{nam}, it has been illustrated that MMG theory minimally coupled with electromagnetic Maxwell-Chern-Simons Lagrangian admits the warped $ AdS_{3}$ type black hole solutions by interestingly considering a polynomial ansatz for the electromagnetic potential. 

In this work, by using the exterior algebra notation, the matter coupling in MMG is studied by first considering that matter Lagrangian depends on the metric co-frame 1-forms together with the matter fields and next concentrating on the case where the matter Lagrangian depends on the connection field as well. It should also be noted that within the same formalism, MMG has been worked out in the works \cite{baykal,dereli_1,dereli_2,dereli_6} but without any consideration of the matter coupling. 
The first part of the study is in fact rephrasing of the work \cite{arvanitakis_2} by using the exterior algebra formalism. As we will explicitly show in the present work, the investigation of the matter coupling in the language of differential forms has similar outcomes where in that case the MMG equation involves source 2-form that contains covariant derivative of stress-energy 2-form as well as quadratic forms of matter stress-energy 2-forms analogous to work \cite{arvanitakis_2}. 
On the other hand, when the matter Lagrangian involves terms that are connection-dependent, the analysis of the theory becomes much more compelling. 
Whether the matter coupling is connection-independent or connection-dependent, to obtain the matter-coupled MMG field equation and the resulting source term, the auxiliary field which is introduced in the Lagrangian of MMG theory should be solved analytically. For the case where the matter coupling is connection-independent, a unique analytical solution for the auxiliary fields can be obtained by using the equations coming from the variation of total Lagrangian with respect to co-frame and connection fields. However, if one considers connection-dependent matter coupling, the task becomes more challenging.        
In that case, to obtain the exact solutions for the auxiliary field, one is enforced to consider some specific matter Lagrangians involving connection terms that are at least of the first order. Within this consideration, our aim in this work is to obtain source 2-forms and get the consistency relations in both cases in the language of differential forms. 
In the first part of the paper, we concentrate on minimally coupled matter Lagrangian that does not depend on spacetime connection explicitly. In terms of exterior forms, we construct the source 2-forms for minimally matter-coupled MMG equation. We will also derive the consistency relation in terms of differential forms in order that the matter-coupled MMG field equation be consistent on-shell which is also known as third way consistency \cite{bergshoeff_3}. In the second part of the work, we will consider a general connection-dependent matter Lagrangian and discuss for what circumstances the exact solution of the auxiliary field can be possible. Finally we will particularly concentrate on the Dirac Lagrangian where it involves first order terms in the connection. We will see that by using the field equations obtained from spinor-matter coupled MMG Lagrangian, the auxiliary field can be analytically solved and as a result one gets MMG field equation with coefficients that remarkably depend on spinor fields. By performing necessary calculations, we obtain related source term in terms of differential forms.
We will also derive the consistency relation for spinor-matter coupled MMG equation.

For the organization of the paper, in Section 2, we examine connection-independent matter-coupling in MMG theory. In Section 3, we make an investigation of connection-dependent matter coupling where we specifically concentrate on spinor-matter coupling. Finally, we close with some discussions and comments.

\section{ Connection-independent matter coupling in MMG }
\label{section_1}

In this part, using the language of differential forms, we present the Lagrangian 3-form (or the action density 3-form) of MMG. Then, we consider the minimal matter coupling of MMG first assuming that the matter Lagrangian depends only on co-frame 1-forms but not on the connection. By variational principle, we obtain the field equations by making independent variations with respect to gravitational field variables and auxiliary field.    
We first point out that the basic field ingredients of the MMG Lagrangian are co-frame 1-forms $ e^{a} $, in terms of which the spacetime metric can be decomposed as $ {\mbox{\boldmath ${g}$}}=\eta_{ab}\, e^{a} \otimes
e^{b}$ with $\eta_{ab} = diag ( - + + )\,$ , connection 1-forms $ \omega^{a}\,_{b} $ and auxiliary 1-form field 
$ \lambda^{a} $. Recall that $ e^{a} $ and $ \omega^{a}\,_{b} $ are gravitational field variables. 
Now, the action of minimal massive 3d gravity (without matter coupling) can be expressed in terms of differential forms as
\begin{equation}
I_{MMG} = \int_{N} \, {\cal L}_{MMG}
\label{eqn_1_1}
\end{equation}
where the Lagrangian 3-form takes the form
\begin{eqnarray}
& & {\cal L}_{MMG} = - \frac{\sigma}{2}  R^{ab} \wedge \ast ( e_{a} \wedge e_{b} ) + \Lambda \ast 1    \nonumber \\ 
& & + \frac{1}{2 \mu} \left( \omega^{a}\,_{b} \wedge d \omega^{b}\,_{a} + \frac{2}{3} \omega^{a}\,_{b} \wedge \omega^{b}\,_{c} \wedge \omega^{c}\,_{a} \right) + \lambda_{a} \wedge T^{a} \nonumber \\ 
& & + \frac{\alpha}{2} \lambda_{a} \wedge \lambda_{b} \wedge \ast ( e^{a} \wedge e^{b} ) 
\label{eqn_1_2}
\end{eqnarray}
Here, $ \Lambda $ is cosmological constant while $ \mu $ denotes mass parameter of TMG.  $ \alpha $ and $ \sigma $ are arbitrary dimensionless parameters that may take either sign.\footnote{Note that in \cite{bergshoeff_3}, $ \sigma $ was assumed to satisfy the relation $ \sigma^{2} = 1 $. In this work, we keep it arbitrary as done in \cite{arvanitakis_2}.} For the case where $ \alpha = 0 $, the Lagrangian describes the action density of TMG such that for this case, auxiliary 1-form field $ \lambda_{a} $ can be identified as the Lagrange multiplier 1-form that enforces the zero-torsion constraint for TMG action.
$\ast$ denotes the Hodge operator which maps a $ p$-form to $ (3-p)$-form in 3 dimensions. In addition, oriented $3$-dimensional volume element can be written $ \ast 1 = e^{0} \wedge e^{1} \wedge e^{2} $.
We point that the connection  1-forms $ \omega^{a}\,_{b} $ satisfy the first Cartan structure equations
\begin{equation}
d e^{a} + \omega^{a}\,_{b} \wedge  e^{b} = T^{a}
\label{eqn_1_3}
\end{equation}
where $T^{a}$ denotes torsion 2-forms while the
corresponding curvature 2-forms are obtained from the second Cartan
structure equations
\begin{equation}
R^{a b} = d \omega^{a b} + \omega^{a}\,_{c} \wedge
\omega^{c\,b}\,\, . \label{eqn_1_4}
\end{equation}
We also consider that the spacetime is metric-compatible which implies that $\omega_{a \,b} = - \omega_{b \,a}\, $.
For future use, we also note that $R=\iota_{a} P^{a}$ is the curvature scalar that can be expressed in terms of Ricci 1-forms $P^{a} = \iota_{b} R^{ba} $ where $\iota_{b}$ is the inner product operator that obeys the identity $\iota_{b} e^{a} = \delta_{b}^{a} $. In addition, we remark that dreibeins $ e^{a}\,_{\mu} $ where $ e^{a} = e^{a}\,_{\mu} d x^{ \mu} $ are assumed to be invertible.  
Now, the action of MMG with minimal matter coupling can be expressed as
\begin{equation}
I = \int_{N} \, ( {\cal L}_{MMG} + {\cal L}_{M} (e, \Phi) )
\end{equation}
where the Lagrangian 3-form $ {\cal L}_{M} $ denotes minimally coupled matter Lagrangian. Note that matter Lagrangian depends only on co-frames and some matter fields but not on the connection of the spacetime, i.e 
\begin{equation}
\frac{ \delta {\cal L}_{M} }{ \delta \omega^{ab} } = 0 \, .
\end{equation} 
Then the field equations can be obtained by making independent variations of the total action with respect to co-frames $e^{a}$, connection 1-forms $\omega^{a}\,_{b}$, the auxiliary field $ \lambda_{a} $ and some matter $p$-form fields $ \Phi_{i} $ where index $ i $ implies that matter Lagrangian can involve more than one matter fields.    
Now, applying the variational principle (under the assumption that the variations of the field variables vanish over 2-dimensional boundary of 3-dimensional manifold $N$ such that surface terms have no contribution to field equations)
\begin{equation}
\delta I = 0
\end{equation}
one obtains the following field equations :
\begin{equation}
- \frac{\sigma}{2} \epsilon_{abc} R^{ab} + \Lambda \ast e_{c} + D \lambda_{c} + \frac{\alpha}{2} \epsilon_{abc} \lambda^{a} \wedge \lambda^{b}  + \tau_{c} [ e, \Phi ] = 0 \, ,
\label{eqn_1_5}
\end{equation}
\begin{equation}
- \frac{\sigma}{2} D ( \ast ( e^{a} \wedge e^{b} ) ) + \frac{1}{ \mu } R^{ba} - \frac{1}{2} ( \lambda^{a} \wedge e^{b} - \lambda^{b} \wedge e^{a} ) = 0 \, ,
\label{eqn_1_6}
\end{equation}
\begin{equation}
T^{a} = -  \alpha \lambda_{b} \wedge \ast ( e^{a} \wedge e^{b} ) \, ,
\label{eqn_1_7}
\end{equation}
where we have used 3-dimensional identities,
\begin{equation}
\ast ( e_{a} \wedge e_{b} ) = \epsilon_{abc} e^{c} \, , \qquad \ast ( e_{a} \wedge e_{b} \wedge e_{c} ) = \epsilon_{abc}  \, ,
\label{eqn_1_8}  
\end{equation}
In addition, we note that $ p $-form matter field equation reads  
\begin{equation}
\frac{ \delta {\cal L}_{M} }{ \delta \Phi_{i} } - (-1)^{p} 
d \left( \frac{ \delta {\cal L}_{M} }{\delta ( d \Phi_{i} )} \right) = 0 \, .
\end{equation}
We further recall that $ D $ denotes the exterior covariant derivative operator acting with respect to connection $ \omega^{a}\,_{b} $ such that 
\begin{equation}
D \lambda_{c} = d \lambda_{c} - \omega^{b}\,_{c} \wedge \lambda_{b} \, .
\end{equation}
Notice that (\ref{eqn_1_5}) is the Einstein field equation obtained from the variation with respect to co-frames $ e^{a} $ where $ \tau_{c} $ denotes matter stress-energy 2-form defined as
\begin{equation}
\tau_{c} = \frac{ \delta {\cal L}_{M}}{ \delta e^{c}} \, . 
\end{equation}
Note that stress-energy 2-form $ \tau_{c} [ e , \Phi ] $ is independent of the connection.  
(\ref{eqn_1_6}) is the equation obtained from variation with respect to connection $ \omega^{a}\,_{b} $ while (\ref{eqn_1_7}) is the equation obtained from the variation of the action with respect to auxiliary 1-form $ \lambda^{a} $ which determines the torsion $T^{a}$ in terms of the auxiliary 1-form field $ \lambda_{b} $. Furthermore, from the relation
\begin{equation}
K^{a}\,_{b} \wedge e^{b} = T^{a} \, ,
\label{eqn_1_10}
\end{equation}
one can obtain the contorsion one forms $K^{a}\,_{b} $ in the form
\begin{equation}
K^{a}\,_{b} = \alpha \epsilon^{a}\,_{bc} \lambda^{c} \, . 
\label{eqn_1_11}
\end{equation}
Recall that the equations  (\ref{eqn_1_5}) and (\ref{eqn_1_6}) involve connection $ \omega^{a}\,_{b}$ with torsion. In addition, the curvature term $R^{ba}$ involves torsion-dependent terms through the contorsion relation when expressed in terms of torsion-free connection. Now, we remark that In order to obtain an algebraic solution for auxiliary field $ \lambda^{a} $, one needs to express all the field equations in terms of torsion-free connection. To this end, we decompose the connection one-form field $ \omega^{a}\,_{b} $ in the form 
\begin{equation}
\omega^{a}\,_{b} = \bar{\omega}^{a}\,_{b} + K^{a}\,_{b} 
\label{eqn_1_12}
\end{equation}
that involves torsion-free (Levi-Civita) connections $ \bar{\omega}^{a}\,_{b} $ and contorsion one-forms $ K^{a}\,_{b} $. Using Cartan structure equation (\ref{eqn_1_4}), one can also express curvature two forms $R^{ab} $ 
\begin{equation}
R^{ab} = \bar{R}^{ab} + \bar{D} K^{ab} + K^{a}\,_{c} \wedge K^{cb}
\label{eqn_1_13}
\end{equation}
in terms of curvature two forms $ \bar{R}^{ab} $ of torsion-free connection $ \bar{\omega}^{a}\,_{b} $ and contorsion. Here $ \bar{D} $ denotes the exterior covariant derivative operator acting with respect to torsion-free connection as
\begin{equation}
\bar{D} K^{ab} = d K^{ab} + \bar{\omega}^{a}\,_{c} \wedge K^{cb} + K^{a}\,_{c} \wedge \bar{\omega}^{cb} 
\label{eqn_1_14}
\end{equation} 
Next, we also point out that 
\begin{equation}
D ( \ast  ( e^{a} \wedge e^{b} ) )  = T^{c} \wedge \ast ( e^{a} \wedge e^{b} \wedge e_{c} ) 
\label{eqn_1_15}  
\end{equation}
such that with the substitution of $ T^{a} $, it yields
\begin{equation}
D ( \ast ( e^{a} \wedge e^{b} ) ) = \alpha ( \lambda^{a} \wedge e^{b} - \lambda^{b} \wedge e^{a} ) \, .
\label{eqn_1_16}
\end{equation}
Then using (\ref{eqn_1_13}) together with (\ref{eqn_1_11}), one can express curvature 2-form $ R^{ab} $ of connection with torsion in terms of curvature 2-form $ \bar{R}^{ab} $ of torsion-free connection and auxiliary 1-form $ \lambda^{a} $. A direct computation yields
\begin{equation}
R^{ab} = \bar{R}^{ab} +  \alpha \epsilon^{abc} \bar{D} \lambda_{c} + \alpha^{2} \lambda^{a} \wedge \lambda^{b}  \, . 
\label{eqn_1_17}
\end{equation} 
Then substituting (\ref{eqn_1_17}) into equations (\ref{eqn_1_5}) and (\ref{eqn_1_6}) , one finally obtains following field equations expressed in terms of torsion-free connection 
\begin{equation}
- \frac{\sigma }{2} \epsilon_{abc} \bar{R}^{ab} + ( \alpha \sigma + 1 ) \bar{D} \lambda_{c} - \frac{1}{2} \alpha ( \alpha \sigma + 1 ) \epsilon_{abc} \lambda^{a} \wedge \lambda^{b} + \Lambda \ast e_{c} + \tau_{c} = 0 
\label{eqn_1_18}  
\end{equation}
and
\begin{equation}
- \frac{1}{2} ( \alpha \sigma + 1 ) ( \lambda^{a} \wedge e^{b} - \lambda^{b} \wedge e^{a} ) + \frac{1}{\mu} ( \bar{R}^{ba} + \alpha \, \epsilon^{bac} \bar{D} \lambda_{c} + \alpha^{2} \lambda^{b} \wedge \lambda^{a} ) = 0 
\label{eqn_1_19}
\end{equation}
where it is also assumed that $ \alpha \sigma + 1 \neq 0 $.  
Furthermore, it is obvious that to obtain the solution for the auxiliary field $ \lambda^{a} $, one should use the equation coming from the variation with respect to connections. It is seen that the connection field equation (\ref{eqn_1_19}) expressed in terms of torsion-free connection involves terms which are quadratic in auxiliary 1-form $ \lambda^{a} $ as well as covariant derivative of $ \lambda^{a} $. To solve for $ \lambda^{a} $, one should eliminate terms $ \alpha \epsilon^{bac} \bar{D} \lambda_{c} $ and $ \alpha^{2} \lambda^{b} \wedge \lambda^{a} $. We note that these terms can be eliminated by using Einstein field equation (\ref{eqn_1_18}). After some algebraic work, one obtains
\begin{equation}
\lambda^{a} \wedge e^{b} - \lambda^{b} \wedge e^{a} = \frac{2}{\mu ( \alpha \sigma + 1 )^{2} } ( \bar{R}^{ba} - \alpha \epsilon^{bac} \tau_{c} + \alpha \Lambda e^{b} \wedge e^{a} ) 
\label{eqn_1_20}
\end{equation}  
which can be solved by the method illustrated in Appendix section. By employing exterior algebra manipulations, the solution can be obtained as
\begin{equation}
\lambda^{a} = - \frac{2}{ \mu ( \alpha \sigma + 1 )^{2} } \left( \bar{Y}^{a} + \frac{ 1 }{2} \alpha \Lambda e^{a} + \alpha \epsilon^{abc} \iota_{b} \tau_{c} - \frac{1}{4} \alpha \epsilon^{npc} \iota_{n} ( \iota_{p} \tau_{c} ) e^{a} \right) 
\label{eqn_1_21} 
\end{equation}
where 
\begin{equation}
\bar{Y}^{a} = \bar{P}^{a} - \frac{ 1 }{4} \bar{R} e^{a}
\label{eqn_1_22}
\end{equation}
denotes Schouten 1-forms expressed in terms of torsion-free Ricci 1-forms $ \bar{P}^{a} $ and the curvature scalar $ \bar{R} $. Furthermore it would be more practical and useful to obtain a more simplified form of $ \lambda^{a}$. For that, we express stress-energy 2-forms $ \tau_{c} $ in the form
\begin{equation}
\tau_{c} = \tau_{cn} \ast e^{n} \, ,
\end{equation}
where $ \tau_{cn} $ gives the components of stress-energy tensor with respect to orthonormal frame.
Next, a direct calculation gives 
\begin{equation}
\epsilon^{abc} \iota_{b} \tau_{c} - \frac{1}{4}  \epsilon^{npc} \iota_{n} ( \iota_{p} \tau_{c} ) e^{a} = - \tau_{n}\,^{a} e^{n} + \frac{ 1 }{2} \tau e^{a}  
\end{equation}
where $ \tau $ denotes trace of stress-energy tensor that can be calculated from the relation $ \tau \ast 1 = e^{c} \wedge \tau_{c} $. At this stage, one can further define a new stress-energy 2-form as
\begin{equation}
\hat{\tau}_{c} = \tau_{c} - \frac{ 1 }{2} \tau \ast e_{c}
\end{equation}
leading to expression
\begin{equation}
- \tau_{n}\,^{a} e^{n} + \frac{ 1 }{2} \tau e^{a}  = \ast \hat{\tau}^{a} \, . 
\end{equation}
As a result, auxiliary 1-form $ \lambda^{a} $ can be expressed in the following simplified form that reads
\begin{equation}
\lambda^{a} = - \frac{2}{ \mu ( \alpha \sigma + 1 )^{2} } \left( \bar{Y}^{a} + \frac{ 1 }{2} \alpha \Lambda e^{a} + \alpha \ast \hat{\tau}^{a} \right) \, . 
\label{auxiliary}
\end{equation} 
Recall that the expression (\ref{auxiliary}) expressed in terms of differential forms is analogous to the expression for the auxiliary field $h_{\mu \nu} $ (the expression (3.8)) presented in \cite{arvanitakis_2}. 
In addition, note that the equation (\ref{eqn_1_20}) is used to get the analytical solution of auxiliary 1-form. Then, if one substitutes (\ref{auxiliary}) into Einstein field equation (\ref{eqn_1_18}) expressed in terms of torsion-free connection, one obtains the field equation of minimally matter-coupled MMG theory where the matter Lagrangian is independent of connection. 
After straightforward calculations, one gets 
\begin{equation}
\alpha_{1} \epsilon_{abc} \bar{R}^{ab} + \alpha_{2} \bar{C}_{c} + \alpha_{3} \ast e_{c} + \frac{ 1 }{2} \alpha_{4} \epsilon_{abc} \bar{Y}^{a} \wedge \bar{Y}^{b} + \tilde{\tau}_{c}= 0
\label{eqn_1_23}
\end{equation}
where 
\begin{equation}
\bar{C}_{c} = \bar{D} \bar{Y}_{c}
\label{eqn_1_24}
\end{equation}
denotes Cotton 2-forms with respect to torsion-free connection and the coefficients read
\begin{eqnarray}
\alpha_{1} = - \frac{ 1 }{2} \left( \sigma + \frac{ 2 \alpha^{2} \Lambda }{ \mu^{2} ( \alpha \sigma + 1 )^{3}} \right) \qquad \alpha_{2} = - \frac{2}{ \mu ( \alpha \sigma + 1 ) } \nonumber \\
\alpha_{3} = \Lambda - \frac{ \alpha^{3} \Lambda^{2} }{ \mu^{2} ( \alpha \sigma + 1 )^{3} } \qquad \alpha_{4} = - \frac{ 4 \alpha }{ \mu^{2} ( \alpha \sigma + 1 )^{3} } \, .
\label{eqn_1_25}
\end{eqnarray}
Also remark that while obtaining MMG field equation (\ref{eqn_1_23}), we use the identity 
\begin{equation}
\bar{R}^{ab} = \bar{Y}^{a} \wedge e^{b} - \bar{Y}^{b} \wedge e^{a} \, .
\label{eqn_1_26}
\end{equation}
Note that remarkable feature of matter-coupled MMG equation is that it involves the source term $ \tilde{\tau}_{c} $ which can be expressed explicitly as
\begin{eqnarray}
\tilde{\tau}_{c} &=& \left( 1 - \frac{ 2 \alpha^{3} \Lambda }{ \mu^{2} ( \alpha \sigma + 1 )^{3} }  \right) \tau_{c} - \frac{ 2 \alpha^{2} }{\mu^{2} ( \alpha \sigma + 1 )^{3} } \epsilon_{abc} ( \bar{Y}^{a} \wedge \ast \hat{\tau}^{b} - \bar{Y}^{b} \wedge \ast \hat{\tau}^{a} ) \nonumber \\
& & - \frac{ 2 \alpha }{ \mu ( \alpha \sigma + 1 ) } \bar{D} ( \ast \hat{\tau}_{c} ) - \frac{ 2 \alpha^{3} }{ \mu^{2} ( \alpha \sigma + 1 )^{3} } \epsilon_{abc} \ast \hat{\tau}^{a} \wedge \ast \hat{\tau}^{b} \, .
\label{source_term}
\end{eqnarray} 
Note that the expression (\ref{source_term}) given in terms of exterior forms is analogous to the expression (3.12) in tensorial notation introduced in \cite{arvanitakis_2}. Similar to the source term in \cite{arvanitakis_2}, the source 2-form (\ref{source_term}) involves explicitly the terms which are quadratic in stress-energy 2-forms. Clearly, source term also depends on covariant derivative of stress-energy 2-form. 

Next, we examine the consistency of matter-coupled MMG equation. For the consistency check, we first note that Einstein 2-forms $ \bar{G}_{c} $ defined by
\begin{equation}
\bar{G}_{c} = - \frac{ 1 }{2} \bar{R}^{ab} \wedge \ast ( e_{a} \wedge e_{b} \wedge e_{c} ) =  - \frac{ 1 }{2} \epsilon_{abc} \bar{R}^{ab}\
\label{eqn_1_27}
\end{equation} 
and the Cotton 2-forms $ \bar{C}_{c} $ satisfy the Bianchi identities 
\begin{equation}
\bar{D} \bar{G}_{c} = 0 \, , \qquad \bar{D} \bar{C}_{c} = 0
\label{eqn_1_28}
\end{equation}
where one also uses the identities
\begin{equation} 
\bar{D} \bar{R}^{ab} = 0 
\label{eqn_1_29}
\end{equation}
and 
\begin{equation}
\bar{D} \bar{C}_{c} = \bar{D}^{2} \bar{Y}_{c} = \bar{R}_{cb} \wedge \bar{Y}^{b} = 0 \, . 
\label{eqn_1_30}
\end{equation} 
In addition, we point out that $ \bar{D} ( \ast e_{c} ) = 0 $ for a torsion-free connection. 
Now one can define $ \bar{J} $-form (similar to $ J $-tensor defined in \cite{bergshoeff_3}) 
\begin{equation}
\bar{J}_{c} = \frac{ 1 }{2} \epsilon_{abc} \bar{Y}^{a} \wedge \bar{Y}^{b} \, ,
\label{eqn_1_31}
\end{equation}
such that in terms of Ricci 1-forms and curvature scalar it reads
\begin{equation}
\bar{J}_{c} = \frac{ 1 }{2} \left( \epsilon_{abc} \bar{P}^{a} \wedge \bar{P}^{b} + \frac{ 1 }{2} \bar{R} \ast \bar{P}_{c} - \frac{3}{8} \bar{R}^{2} \ast e_{c} \right) \, .
\end{equation}
An interesting result comes out from the exterior multiplication of $\bar{J}_{c}$ by $ e^{c}$. The calculation yields
\begin{equation}
\bar{J}_{c} \wedge e^{c} = \frac{ 1 }{2} \left( \frac{3}{8} \bar{R}^{2} \ast 1 - \bar{P}_{a} \wedge \ast \bar{P}^{a} \right)
\end{equation}  
that can also be identified as trace of $J$-form if one expresses $\bar{J}_{c}$ in the form $ \bar{J}_{c} = \bar{J}_{cp} \ast e^{p} $ (similar to stress-energy 2-form). It is seen that the result of such a multiplication is surprisingly identical to Lagrangian 3-form of NMG (or $K$-tensor also mentioned in \cite{arvanitakis_1}). This remarkable coincidence leads to quadratic equations for cosmological constant of maximally-symmetric spacetimes for both NMG and MMG theories without matter-coupling. It should also be noted that for both NMG and MMG, there exists a unique maximally-symmetric vacuum (at the so-called merger point) for a critical value of the cosmological parameter $\Lambda$. Now, in the exterior algebra formalism, let us first examine the consistency of vacuum MMG equation without any matter-coupling. Then, the consistency requires that (when acting operator $ \bar{D}$)
\begin{equation}
\bar{D} \bar{J}_{c} = 0 \, .
\end{equation}
On the other hand, from the definition (\ref{eqn_1_31}), a straightforward calculation yields
\begin{equation}
\bar{D} \bar{J}_{c} = \epsilon_{abc} \bar{C}^{a} \wedge \bar{Y}^{b}
\end{equation}
which is not identically zero. However, if one eliminates Cotton 2-form from the MMG field equation (by taking $ \tau_{c} = 0 $ and $ \tilde{\tau}_{c} = 0 $ in absence of matter coupling), one remarkably obtains 
\begin{equation}
\bar{D} \bar{J}_{c} = 0 
\end{equation}
owing to the identities
\begin{eqnarray}
\epsilon_{abc} \epsilon^{a}\,_{de} \bar{R}^{de} \wedge \bar{Y}^{b} = 2 ( \bar{Y}_{c} \wedge e_{b} - \bar{Y}_{b} \wedge e_{c} ) \wedge \bar{Y}^{b} = 0 \nonumber \\
\epsilon_{abc} \bar{Y}^{b} \wedge \ast e^{a} = 0  
\label{eqn_1_32} \\
\epsilon_{abc} \epsilon^{a}\,_{de} \bar{Y}^{d} \wedge \bar{Y}^{e} \wedge \bar{Y}^{b} = 0 \nonumber
\end{eqnarray}  
that hold for torsion-free spacetimes. As is also mentioned in \cite{bergshoeff_3} and \cite{arvanitakis_2}, this implies that the vacuum MMG equation is consistent on shell. 
Now, for the consistency of matter-coupled MMG equation, one similarly acts the exterior covariant derivative operator on matter-coupled MMG equation. Using the arguments mentioned above, the consistency requires that 
\begin{equation}
\bar{D} \bar{J}_{c} + \bar{D} \tilde{\tau}_{c} = 0
\end{equation}
or equivalently
\begin{equation}
\epsilon_{abc} \bar{C}^{a} \wedge \bar{Y}^{b} + \bar{D} \tilde{\tau}_{c} = 0 \, .
\end{equation}
As before, eliminating Cotton tensor from matter-coupled MMG equation, one obtains 
\begin{equation}
\bar{D} \tilde{\tau}_{c} = \frac{ \alpha_{4} }{ \alpha_{2} } \epsilon_{abc} \tilde{ \tau }^{a} \wedge \bar{Y}^{b} \, .
\label{consistency}
\end{equation}
where again the identities (\ref{eqn_1_32}) are used. Recall that the expression (\ref{consistency}) expressed in terms of differential forms is analogous to expression (4.15) in \cite{arvanitakis_2}.   
Note that the consistency of MMG equation (\ref{eqn_1_23}) with connection-independent matter coupling requires the source term (\ref{source_term}) to satisfy the relation (\ref{consistency}).
In order to illustrate that the source term satisfies the consistency relation, one simply calculates the left and right hand sides of this relation and checks their equality. In \cite{arvanitakis_2}, the equality of left and right hand sides of the consistency equation has been illustrated by simply taking cosmological term to be equal to zero (which corresponds to taking $ \alpha_{3} = 0 $ in the present work). In fact, it can be explicitly proven that the source term (\ref{source_term}) satisfies the consistency equation (\ref{consistency}) for arbitrary values of the cosmological parameter. To show the equality of both sides of the consistency relation, we first calculate the covariant derivative of the source term for left hand side. Noting that the stress-energy 
2-forms $ \tau_{c} $ are covariantly conserved i.e $ \bar{D} \tau_{c} = 0 $, the left hand side can be calculated as
\begin{eqnarray}
\bar{D} \tilde{\tau}_{c} = - \frac{ 2 \alpha }{ \mu ( \alpha \sigma + 1 ) } \bar{R}_{cb} \wedge \ast \hat{\tau}^{b} \nonumber \\
- \frac{ 4 \alpha^{2} }{ \mu^{2} ( \alpha \sigma + 1 )^{3} } \, \epsilon_{abc} ( \bar{C}^{a} \wedge \ast \hat{\tau}^{b} - \bar{Y}^{a} \wedge \bar{D} ( \ast \hat{\tau}^{b} ) ) \nonumber  \\
- \frac{ 4 \alpha^{3} }{ \mu^{2} ( \alpha \sigma + 1 )^{3} } \, \epsilon_{abc} \bar{D} ( \ast \hat{\tau}^{a} ) \wedge \ast \hat{\tau}^{b}
\label{consistency_check_1}
\end{eqnarray}
where we have used
\begin{equation}
\bar{D}^{2} ( \ast \hat{\tau}_{c} ) = \bar{R}_{cb} \wedge \ast \hat{\tau}^{b} \, .
\end{equation} 
Furthermore if one eliminates the Cotton 2-form by using MMG field equation (\ref{eqn_1_23}) and makes some straightforward calculations, one finally obtains the following expression for the left hand side 
\begin{eqnarray}
\bar{D} \tilde{\tau}_{c} = \left( \frac{ 8 \alpha^{2} \, \alpha_{1} }{ \mu^{2} ( \alpha \sigma + 1 )^{3} \, \alpha_{2} } - \frac{ 2 \alpha }{ \mu ( \alpha \sigma + 1 ) } \right) \bar{R}_{cb} \wedge \ast \hat{\tau}^{b} \nonumber \\
- \frac{ 4 \alpha^{2} \, \alpha_{4} }{ \mu^{2} ( \alpha \sigma + 1 )^{3} \, \alpha_{2} } \bar{Y}_{c} \wedge \ast \hat{\tau}_{b} \wedge \bar{Y}^{b} + \frac{ 16 \alpha^{4} }{ \mu^{4} ( \alpha \sigma + 1 )^{6} \, \alpha_{2} } \ast \hat{\tau}_{c} \wedge \ast \hat{\tau}_{b} \wedge \bar{Y}^{b} \nonumber \\
- \frac{ 4 \alpha^{2} }{ \mu^{2} ( \alpha \sigma + 1 )^{3} } \epsilon_{abc} \, \bar{D} ( \ast \hat{\tau}^{a} ) \wedge \bar{Y}^{b} \, .
\label{consistency_check_2}
\end{eqnarray}
For the right hand side, we simply evaluate $ \epsilon_{abc} \, \tilde{\tau}^{a} \wedge \bar{Y}^{b} $ and compare both sides. Performing the required calculations by also making use of curvature-Schouten identity (\ref{eqn_1_26}) and the identities 
\begin{equation}
\bar{Y}_{c} \wedge e_{b} \wedge \ast \hat{\tau}^{b} = 0 
\label{identity_1}
\end{equation}
and 
\begin{equation}
\epsilon_{abc} \tau^{a} \wedge \bar{Y}^{b} = \bar{Y}^{b} \wedge e_{c} \wedge \ast \tau_{b} \, ,
\label{identity_2}
\end{equation} 
the left and the right hand sides of the consistency relation (\ref{consistency}) can be proven to be equal for arbitrary values of cosmological term $ \alpha_{3} $ and therefore for arbitrary values of the cosmological parameter $ \Lambda $.

\vspace{0.4cm}

\noindent {\bf Some connection-independent matter Lagrangian 3-forms :}

\vspace{0.4cm}

\noindent Now to illustrate the method that we use, we consider some sample matter Lagrangians in exterior algebra notation that are minimally coupled to MMG. We note that the matter Lagrangians considered do not depend on connection 1-form fields $ \omega^{a}\,_{b} $.

\vspace{0.4cm}

\noindent {\bf i.} Maxwell Chern-Simons Lagrangian 3-form :   

\vspace{0.4cm}
\noindent  Our first sample Lagrangian 3-form is the Maxwell Chern-Simons Lagrangian that explicitly reads
\begin{equation}
{\cal L}_{m} [e^{a},  A] = - \frac{ 1 }{2} F \wedge \ast F - \frac{ 1 }{2} m_{e} A \wedge F
\end{equation}
where $ A $ denotes electromagnetic potential 1-form field while $ F $ describes electromagnetic 2-form field such that $ F = d A $. $ m_{e} $ corresponds to mass parameter associated with Maxwell field. The variation of the Maxwell-Chern-Simons Lagrangian 3-form with respect to co-frame field $ e^{a} $ gives the stress-energy 2-forms
\begin{equation}
\tau_{c} = \frac{ \delta {\cal L}_m [ e^{a} , A ] }{ \delta e^{c} } = \frac{ 1 }{2} ( \iota_{c} F \wedge \ast F - F \wedge \iota_{c} ( \ast F ) ) 
\end{equation}
The trace of stress-energy 2-form can be calculated from
\begin{equation}
\tau  \ast 1 = e^{c} \wedge \tau_{c} = \frac{ 1 }{2} F \wedge \ast F
\end{equation}
where we use the identity 
\begin{equation}
e^{c} \wedge \iota_{c} \Omega = p \, \Omega
\label{identity}
\end{equation}
for any $p$-form field $ \Omega $. Then the auxiliary 1-form field $ \lambda^{a} $ can be calculated from (\ref{auxiliary}).
Also we note that variation of Maxwell Chern-Simons Lagrangian with respect to potential 1-form $ A $ produces field equation
\begin{equation}
d ( \ast F ) + m_{e} F = 0 \, .
\end{equation}

\vspace{0.4cm}
\noindent {\bf ii.} Scalar matter Lagrangian 3-form with a potential term : 

\vspace{0.4cm}
\noindent Our next sample Lagrangian 3-form is the scalar matter Lagrangian that can be expressed in the form
\begin{equation}
{\cal L}_{m} [e^{a}, \phi] = - \frac{ 1 }{2} d \phi \wedge \ast d \phi - U(\phi) \ast 1
\end{equation}   
that includes the potential term $ U(\phi) $ as well. In this case, the variation of scalar matter Lagrangian with respect to co-frame $ e^{a} $ yields stress-energy 2-forms
\begin{equation}
\tau_{c} = \frac{ \delta {\cal L}_{m} [ e^{a} , \phi] }{\delta e^{c} } = \frac{ 1 }{2} ( \iota_{c} d \phi \wedge \ast d \phi + d \phi \wedge \iota_{c} ( \ast d \phi ) ) - U(\phi) \ast e_{c} \, .
\end{equation}
The trace of stress-energy 2-form can be obtained from the expression
\begin{equation}
\tau \ast 1 = e^{c} \wedge \tau_{c}  = - \frac{ 1 }{2} d \phi \wedge \ast d \phi - 3 U(\phi) \ast 1 
\end{equation}
where again the identity (\ref{identity}) has been used. Then the auxiliary 1-form $ \lambda^{a} $ can be calculated from (\ref{auxiliary}). In addition, by making variation of matter Lagrangian with respect to scalar field $ \phi $ produces matter field equation
\begin{equation}
d ( \ast d \phi ) = U^{\prime} ( \phi) \ast 1
\end{equation}
where $ U^{\prime} (\phi ) $ denotes the ordinary derivative of potential term with respect to scalar field $ \phi $ (i.e $ U^{\prime} ( \phi ) = \frac{d U}{d  \phi } $).

\section{ Connection-dependent matter coupling in MMG  }   
\label{section_2}   

\noindent  In this section, the matter Lagrangian is considered to be dependent on co-frames and connection 1-forms in general. As in the previous section, we concentrate on the minimal coupling. Then the corresponding field equations can be obtained by considering the independent variations of the total action
\begin{equation}
I = \int_{N} ( {\cal L}_{MMG} + {\cal L}_M (e,\omega, \Phi) )
\end{equation}
with respect to co-frames $ e^{a} $, the connections $ \omega^{a}\,_{b} $, the auxiliary 1-form field $ \lambda_{a} $ and some matter fields. Now the co-frame and the connection variations of the total action yield the field equations 
\begin{equation}
- \frac{\sigma}{2} \epsilon_{abc} R^{ab} + \Lambda \ast e_{c} + D \lambda_{c} + \frac{\alpha}{2} \epsilon_{abc} \lambda^{a} \wedge \lambda^{b}  + \tau_{c} [ e, \omega , \Phi ] = 0 \, ,
\label{einstein_2_1}
\end{equation}
and
\begin{equation}
- \frac{\sigma}{2} D ( \ast ( e^{a} \wedge e^{b} ) ) + \frac{1}{ \mu } R^{ba} - \frac{1}{2} ( \lambda^{a} \wedge e^{b} - \lambda^{b} \wedge e^{a} ) + \Omega^{ab} [ e, \omega , \Phi ] = 0 \, 
\label{connection_2_1}
\end{equation}
respectively where 
\begin{equation}
\Omega_{ab} = \frac{ \delta {\cal L}_{M} }{ \delta \omega^{ab} } \, .
\end{equation}
Note that $ \Omega^{ab}$ is identified as hyper-momentum 2-forms coming from the connection variation of the matter Lagrangian. Also recall that for this case, the stress-energy and the hyper-momentum 2-forms are connection-dependent in general. On the other hand, $ \lambda_{a} $ variation produces the torsion equation (\ref{eqn_1_7}) as before. As in the connection-independent part, to find an analytical solution for the auxiliary field, it would be convenient to express the field equations (\ref{einstein_2_1}) and (\ref{connection_2_1}) in terms of torsion-free connection terms. Therefore considering the decomposition of connection 1-forms and the $\lambda $-dependancy of the contorsion (the expressions (\ref{eqn_1_11}) and (\ref{eqn_1_12})), one can also express the stress-energy and hyper-momentum 2-forms in the convenient forms  
\begin{equation}
\tau_{c} [e,\omega,\Phi] = \bar{\tau}_{c} [e,\bar{\omega},\Phi] + \Gamma_{c} [e,\lambda,\Phi]
\end{equation}
and
\begin{equation}
\Omega^{ab} [e,\omega,\Phi] = \bar{\Omega}^{ab} [e,\bar{\omega}, \Phi] + \tilde{\Omega}^{ab} [e,\lambda,\Phi]
\end{equation}
respectively where the barred terms denote stress-energy and hyper-momentum 2-forms expressed in terms of torsion-free connection $ \bar{\omega} $. Then the resulting field equations can be written as 
\begin{equation}
- \frac{\sigma }{2} \epsilon_{abc} \bar{R}^{ab} + ( \alpha \sigma + 1 ) \bar{D} \lambda_{c} - \frac{1}{2} \alpha ( \alpha \sigma + 1 ) \epsilon_{abc} \lambda^{a} \wedge \lambda^{b} + \Lambda \ast e_{c} + \bar{\tau}_{c} + \Gamma_{c} = 0 
\label{einstein_2_2}  
\end{equation}
\begin{equation}
- \frac{1}{2} ( \alpha \sigma + 1 ) ( \lambda^{a} \wedge e^{b} - \lambda^{b} \wedge e^{a} ) + \frac{1}{\mu} ( \bar{R}^{ba} + \alpha \, \epsilon^{bac} \bar{D} \lambda_{c} + \alpha^{2} \lambda^{b} \wedge \lambda^{a} ) + \bar{\Omega}^{ab} + \tilde{\Omega}^{ab} = 0 \, .
\label{connection_2_2}
\end{equation}
Next using (\ref{einstein_2_2}), the equation (\ref{connection_2_2}) can be put into the form
\begin{eqnarray}
- \frac{ 1 }{2} ( \alpha \sigma + 1 ) ( \lambda^{a} \wedge e^{b} - \lambda^{b} \wedge e^{a} ) \nonumber \\ 
+ \frac{ 1 }{\mu ( \alpha \sigma + 1 ) } ( \bar{R}^{ba} - \alpha \epsilon^{bac} (\bar{\tau}_{c} + \Gamma_{c} ) + \alpha \Lambda e^{b} \wedge e^{a} ) + \bar{\Omega}^{ab} + \tilde{\Omega}^{ab} = 0 \, .
\label{connection_2_3}
\end{eqnarray}   

\noindent Now obtaining an algebraic solution for $ \lambda^{a} $ certainly depends on the forms of the terms $ \Gamma_{c} $ and $ \tilde{\Omega}^{ab} $. In general, these terms can involve second or higher order terms in $ \lambda_{a} $ as well as the covariant derivatives of the auxiliary field that make (\ref{connection_2_3}) a highly non-linear equation in $ \lambda_{a} $. Therefore an algebraic solution for auxiliary field may not be possible. To illustrate it better, let us consider the following matter coupling in MMG described by matter Lagrangian in the form 
\begin{equation}
{\cal L} = {\cal L}_{MMG} + \frac{ 1}{2} \Phi^{ab} R_{ac} \wedge \ast R^{c}\,_{b} + \frac{ 1 }{2} \beta \, D \Phi_{ ab } \wedge \ast D \Phi^{ ab }
\label{matter_lagrangian}
\end{equation}
where $ \Phi_{ab} $ is symmetric second order tensor-valued field as introduced in \cite{dereli_5} while $ \beta $ denotes coupling constant. For such a type of matter coupling, if one considers the variations with respect to co-frames and connections and expresses the resulting field equations in terms of Levi-Civita connection, it can obviously be seen that the terms 
$ \Gamma_{c} $ and $ \tilde{\Omega}^{ab} $ involve second and higher order terms in $ \lambda^{a} $ as well as its second order covariant derivatives. For this sample matter Lagrangian, it is not clear that how such terms can be eliminated. Therefore for such cases, an algebraic solution may not be possible. 

At this point, one can claim that if the matter Lagrangian involves terms that are of first order in connection (i.e the terms that are linear in connection), then one can hopefully get an algebraic solution for auxiliary field. For such a form of matter Lagrangian, the term $ \Gamma_{c} $ obviously becomes proportional to auxiliary field while the term $ \tilde{\Omega}^{ab} $ would be independent of $ \lambda^{a} $. This is indeed the case for Dirac spinor matter Lagrangian where it involves the covariant derivatives of spinor fields that are of first order in connection. As we will see in the following, such a connection dependency of the Dirac Lagrangian enables us to calculate the auxiliary 1-forms $ \lambda^{a} $ analytically. Hence in what follows, we will specifically focus on the MMG theory that is minimally coupled to Dirac spinor fields. We notice that we will consider a spinor-matter coupling as minimally as possible. Before proceeding any further, a discussion should also be made on how to couple the spinor fields minimally to gravitational field variables especially to connection field. There exist two alternatives for that coupling. One may either couple the spinor field to a Levi-Civita connection with no torsion or one may consider the spinor-matter coupling to a connection 1-form with non-zero torsion. We prefer the latter due to two main arguments. First, we notice that MMG is a 3d theory involving torsion. Furthermore, if one considers the Einstein-Cartan-Dirac theory described by the Lagrangian 3-form   
\begin{equation}
{\cal L}_{ECD} = - \frac{ 1 }{2 \kappa } R^{ab} \wedge \ast ( e^{a} \wedge e^{b} ) + \Lambda \ast 1 + \frac{i}{2} ( \bar{\psi} \ast e^{a} \gamma_{a} \wedge D \psi - \ast e^{a} \wedge D \bar{\psi} \, \gamma_{a} \, \psi ) - i m \bar{\psi} \psi \ast 1 
\label{einstein_cartan_dirac}
\end{equation}
discussed in \cite{dereli_4}, it can clearly be seen that independent connection variations of (\ref{einstein_cartan_dirac}) produce spacetime torsion expressed in terms of spinor field as is also illustrated in \cite{dereli_4}. 

Motivated by these arguments, it is natural to consider minimal coupling of spinor field to MMG theory by taking the action
\begin{equation}
I = \int_{N} \, ( {\cal L}_{MMG} + {\cal L}_{D} ) 
\label{mmg_dirac}
\end{equation}
where $ {\cal L}_{MMG} $ is given in (\ref{eqn_1_2}) and $ {\cal L}_{D} $ reads
\begin{equation}
{\cal L}_{D} = \frac{i}{2} ( \bar{\psi} \ast e^{a} \gamma_{a} \wedge D \psi - \ast e^{a} \wedge D \bar{\psi} \, \gamma_{a} \, \psi ) - i m \bar{\psi} \psi \ast 1 
\label{eqn_2_2}
\end{equation}
where the Dirac spinor field $ \psi $ is minimally coupled to connection 1-form involving spacetime torsion. Also, recall that in the absence of gravitational Chern-Simons term and the auxiliary field $ \lambda_{a} $, the resulting action (\ref{mmg_dirac}) is exactly the Einstein-Cartan-Dirac action where its independent connection variations lead to non-vanishing torsion. Now before obtaining the field equations, let us also introduce the content of Dirac Lagrangian. Here $ \gamma_{a} $ are generators of $ Cl_{1,2} $ Clifford algebra that obeys the relation
\begin{equation}
\{ \gamma_{a} , \gamma_{b} \} = \gamma_{a} \gamma_{b} + \gamma_{b} \gamma_{a} = 2 \eta_{ab} \, .
\end{equation}   
$ m $ is mass parameter associated with spinor field. Covariant derivative of Dirac field is defined as
\begin{equation}
D \psi = d  \psi + \frac{ 1 }{2} \omega^{cd} \, \sigma_{cd} \psi
\end{equation}
where
\begin{equation}
\sigma_{cd} = \frac{ 1 }{4} [ \gamma_{c} , \gamma_{d} ] = \frac{ 1 }{4} ( \gamma_{c} \gamma_{d} - \gamma_{d} \gamma_{c} ) 
\end{equation}
are generators of algebra of the group $ SO(1,2)$. $ \bar{\psi} $ is Dirac conjugate spinor defined as $ \bar{\psi} = \psi^{\dagger} \gamma_{0} $ whose covariant derivative is given by 
\begin{equation}
D \bar{\psi} =  d \bar{\psi} - \frac{ 1 }{2} \bar{\psi} \sigma_{cd} \, \omega^{cd} \, .
\end{equation}
Choosing the following particular representation 
\begin{equation}
\gamma_{0} = i \left( 
\begin{array}{cc}
1 & 0 \\
0 & - 1 
\end{array} \right) \, , \qquad 
\gamma_{1} = \left(
\begin{array}{cc}
0 & 1 \\
1 & 0 
\end{array} \right) \, , \qquad 
\gamma_{2} = i \left(
\begin{array}{cc}
0 & 1 \\
- 1 & 0 
\end{array} \right)
\label{representation}
\end{equation}
for generators $\gamma_{a} $, one can also prove the identities
\begin{eqnarray}
& & \gamma_{0} \gamma_{a}^{\dagger} \gamma_{0} = \gamma_{a} \, , \qquad \gamma_{a}^{\dagger} \gamma_{0} = - \gamma_{0} \gamma_{a} \, ,  \nonumber \\
& & \sigma_{ab} = \frac{ 1 }{2} \epsilon_{abc} \gamma^{c} \, , \qquad \sigma_{ab} \gamma_{c} + \gamma_{c} \sigma_{ab} = \epsilon_{abc} \, .
\label{gamma_identity}
\end{eqnarray}

\noindent Now, the field equations of spinor-coupled MMG theory can be obtained by considering variations with respect to co-frame fields $ e^{a} $, connections $ \omega^{a}\,_{b} $, auxiliary 1-form $ \lambda^{a} $ and conjugate Dirac spinor field 
$ \bar{\psi} $. Then the variational principle (again under the assumption that field variations vanish over the boundary of 3-dimensional manifold $ N $) 
\begin{equation}
\delta I = 0 
\end{equation}
leads to following field equations: 
\begin{equation}
- \frac{\sigma}{2} \epsilon_{abc} R^{ab} + \Lambda \ast e_{c} + D \lambda_{c} + \frac{\alpha}{2} \epsilon_{abc} \lambda^{a} \wedge \lambda^{b} + \tau^{D}_{c} = 0 \, ,
\label{eqn_2_3}
\end{equation}
\begin{equation}
- \frac{\sigma}{2} D ( \ast ( e^{a} \wedge e^{b} ) ) + \frac{ 1 }{\mu} R^{ba} - \frac{ 1 }{2} ( \lambda^{a} \wedge e^{b} - \lambda^{b} \wedge e^{a} ) - \frac{ 1 }{4} ( i \bar{\psi} \psi ) e^{a} \wedge e^{b} = 0 \, ,
\label{eqn_2_4}
\end{equation}
\begin{equation}
T^{a} = - \alpha \lambda_{b} \wedge \ast ( e^{a} \wedge e^{b} ) \, , 
\label{eqn_2_5}
\end{equation}   
\begin{equation} 
\ast e^{a} \gamma_{a} \wedge D \psi + \frac { 1 }{2} T^{b} \wedge \ast ( e^{a} \wedge e_{b} ) \gamma_{a} \psi - m \psi \ast 1 = 0 \, . \label{eqn_2_6}
\end{equation} 
For later convenience, we also note that the equation of Dirac conjugate spinor field reads 
\begin{equation}
\ast e^{a} \wedge D \bar{\psi}  \gamma_{a} + \frac{ 1 }{2} T^{b} \wedge \ast ( e^{a} \wedge e_{b} ) \bar{\psi} \gamma_{a} + m \bar{\psi} \ast 1 = 0 \, .
\label{eqn_2_7}
\end{equation}
Here, in Einstein equation (\ref{eqn_2_3}), $  \tau^{D}_{c} $ denotes Dirac stress-energy 2-forms defined by
\begin{equation}
\tau^{D}_{ c } = \frac{ \delta {\cal L}_{D}}{ \delta e^{c} } 
\end{equation} 
where it can be obtained explicitly in the form
\begin{equation}
\tau^{D}_{c} = \frac{i}{2} ( \ast( e^{a} \wedge e_{c} ) \wedge \bar{\psi} \gamma_{a} D \psi - \ast( e^{a} \wedge e_{c} )  
\wedge D \bar{\psi} \, \gamma_{a} \psi ) - m ( i \bar{\psi} \psi ) \ast e_{c} \, . 
\label{eqn_2_8}
\end{equation}  
For later use, it would be appropriate to express Dirac stress-energy 2-forms in the form
\begin{equation}
\tau^{D}_{c} = t_{c} - \frac{ 1 }{4} \alpha ( i \bar{\psi} \psi ) \epsilon_{abc} ( \lambda^{a} \wedge e^{b} - \lambda^{b} \wedge e^{a} ) \, .
\label{eqn_2_13}
\end{equation}
where we have used the decomposition of connection 1-form. Also, we have defined $ \bar{\tau}^{D}_{c} = : t_{c} $ as Dirac stress-energy 2-forms expressed in terms of torsion-free connection. It reads explicitly
\begin{equation}
t_{c} =  \frac{ i }{2} \left( \ast ( e^{a} \wedge e_{c} ) \wedge \bar{\psi} \gamma_{a} \bar{D} \psi - \ast ( e^{a} \wedge e_{c} ) \wedge \bar{D} \bar{\psi}  \, \gamma_{a} \psi \right) - m ( i \bar{\psi} \psi ) \ast e_{c} \, . 
\label{eqn_2_14}
\end{equation}  
As in the Section 2, from the torsion expression (\ref{eqn_2_5}), contorsion 1-forms can be obtained in the form
\begin{equation}
K^{a}\,_{b} = \alpha \epsilon^{a}\,_{bc} \lambda^{c} \, .
\end{equation}
Next, we remark that all the field equations have been expressed with respect to a connection with torsion. As in the previous part, in order to calculate auxiliary field $ \lambda^{a} $, it would be convenient to express the field equations with respect to a torsion-free connection. Then using the decomposition of the connection 1-forms in the form
\begin{equation}
\omega^{a}\,_{b} = \bar{\omega}^{a}\,_{b} + K^{a}\,_{b}
\end{equation}
and performing algebraic computations, one obtains the following field equations with respect to torsion-free connection:
\begin{eqnarray}
& &- \frac{\sigma}{2} \epsilon_{abc} \bar{R}^{ab} + ( \alpha \sigma + 1 ) \bar{D} \lambda_{c} + \Lambda \ast e_{c} - \frac{ 1 }{2} \alpha ( \alpha \sigma + 1 ) \epsilon_{abc} \lambda^{a} \wedge \lambda^{b}  + t_{c} \nonumber \\
& & - \frac{ 1 }{4} \alpha ( i \bar{\psi} \psi ) \epsilon_{abc} ( \lambda^{a} \wedge e^{b} - \lambda^{b} \wedge e^{a} ) = 0 \, ,
\label{eqn_2_9}
\end{eqnarray}   
\begin{eqnarray}
& & - \frac{ ( \alpha \sigma + 1 ) }{2} ( \lambda^{a} \wedge e^{b} - \lambda^{b} \wedge e^{a} ) + \frac{ 1 }{\mu} ( \bar{R}^{ba} + \alpha \epsilon^{bac} \bar{D} \lambda_{c} + \alpha^{2} \lambda^{b} \wedge \lambda^{a} ) \nonumber \\
& & + \frac{ 1 }{4} ( i \bar{\psi} \psi ) e^{b} \wedge e^{a} = 0  \, ,
\label{eqn_2_10}
\end{eqnarray}
\begin{equation}
\ast e^{a} \gamma_{a} \wedge \bar{D} \psi - \frac{ 1 }{2} \alpha \lambda_{a} \wedge \ast e^{a} \psi - m \psi \ast 1 = 0 \, .
\label{eqn_2_11}
\end{equation}
Also note that in terms of torsion-free connection, the conjugate spinor field satisfies the field equation 
\begin{equation}
\ast e^{a} \wedge \bar{D} \bar{\psi} \gamma_{a} + \frac{ 1 }{2} \alpha \lambda_{a} \wedge \ast e^{a} \bar{\psi} + m \bar{\psi} \ast 1 = 0 \, .
\label{eqn_2_12}
\end{equation}

\noindent In addition, we point out that equation (\ref{eqn_2_10}) must be used to obtain the algebraic solution of auxiliary 1-form $\lambda^{a}$. As done in connection-independent part, the terms $ \alpha \epsilon^{bac} \bar{D} \lambda_{c} $ and $ \alpha^{2} \lambda^{b} \wedge \lambda^{a} $ should be eliminated. Similarly, these terms can be eliminated by using Einstein field equation (\ref{eqn_2_9}). After simplifications, (\ref{eqn_2_10}) turns into
\begin{equation}
\lambda^{a} \wedge e^{b} - \lambda^{b} \wedge e^{a} = \frac{ 2}{ \mu ( \alpha \sigma + 1 )^{2} \, \theta } \left( \bar{R}^{ba} + \tilde{\Lambda} \, e^{b} \wedge e^{a} - \alpha \epsilon^{bac} t_{c} \right) 
\label{eqn_2_15}
\end{equation}       
where we define the spinor field-dependent functions
\begin{equation}
\tilde{\Lambda} = \alpha \Lambda + \frac{ 1}{4} \mu ( \alpha \sigma + 1) ( i \bar{\psi} \psi )
\label{eqn_2_16}
\end{equation}
and
\begin{equation}
\theta = 1 - \frac{ \alpha^{2} }{ \mu ( \alpha \sigma + 1 )^{2} } ( i \bar{\psi} \psi )  \, .
\label{eqn_2_17}
\end{equation}
Note that the term $ ( i \bar{\psi} \psi ) $ is a real scalar which depends on spacetime coordinates. Therefore, $ \tilde{\Lambda} $ and $ \theta $ are real scalar functions. Now, to obtain the solution of the equation (\ref{eqn_2_15}), we make use of the method presented in Appendix section. Furthermore, as done in the previous section, we express torsion-free Dirac stress-energy 2-form as
\begin{equation}
t_{c} = t_{cp} \ast e^{p}
\label{eqn_2_18}
\end{equation}
and define new Dirac stress-energy form 
\begin{equation}
\hat{t}_{c} = t_{c} - \frac{ 1}{2} t \ast e_{c} \, , 
\label{eqn_2_19}
\end{equation}     
where $ t $ is the trace of Dirac tensor $ t_{cp} $ that can be calculated from the relation $ e^{c} \wedge t_{c} = t \ast 1 $. Finally, the solution of $ \lambda^{a} $ can be obtained as
\begin{equation}
\lambda^{a} = - \frac{ 2 }{ \mu ( \alpha \sigma + 1 )^{2} \, \theta } \left( \bar{Y}^{a} + \frac{ 1 }{2} \tilde{\Lambda} e^{a} + \alpha \ast \hat{t}^{a} \right) \, .
\label{eqn_2_20}
\end{equation}
Now, to get spinor-matter coupled MMG equation, we substitute $ \lambda^{a} $ into Einstein equation (\ref{eqn_2_9}). After necessary computations and simplifications, one gets the spinor-matter coupled MMG equation in the form
\begin{equation}
\beta_{1} \, \epsilon_{abc} \bar{R}^{ab} + \beta_{2} \, \bar{C}_{c} + \beta_{3} \, \ast e_{c} + \frac{ 1 }{2} \beta_{4} \, \epsilon_{abc} \bar{Y}^{a} \wedge \bar{Y}^{b} + \tilde{t}_{c} = 0
\label{eqn_2_21}
\end{equation}  
where
\begin{eqnarray}
& & \beta_{1} = - \frac{ 1 }{2} \left( \sigma + \frac{ 2 \alpha \tilde{\Lambda} }{ \mu^{2} ( \alpha \sigma + 1 )^{3} \theta^{2}} - \frac{ \alpha ( i \bar{\psi} \psi ) }{ \mu ( \alpha \sigma + 1 )^{2} \theta } \right) , \quad \beta_{2} = - \frac{ 2 }{ \mu ( \alpha \sigma + 1 ) \theta } \, , \nonumber \\
& & \beta_{3} = \Lambda - \frac{ \alpha \tilde{\Lambda}^{2} }{ \mu^{2} ( \alpha \sigma + 1 )^{3}  \theta^{2} } + \frac{ \alpha \tilde{\Lambda} ( i \bar{\psi} \psi ) }{ \mu ( \alpha \sigma + 1 )^{2} \theta } \, , \quad \beta_{4} = - \frac{ 4 \alpha }{ \mu^{2} ( \alpha \sigma + 1 )^{3} \theta^{2} } 
\label{eqn_2_22}
\end{eqnarray}
and the source 2-form for spinor-matter coupled MMG reads
\begin{eqnarray}
\tilde{t}_{c} &=& \frac{ 2 }{ \mu ( \alpha \sigma + 1 ) \theta^{2} } \bar{D} \theta \wedge \left( \bar{Y}_{c} + \frac{ 1 }{2} \tilde{\Lambda} e_{c} + \alpha \ast \hat{t}_{c} \right)  \nonumber \\
& & -  \frac{ 1 }{ \mu ( \alpha \sigma + 1 ) \theta } \bar{D} \tilde{\Lambda} \wedge e_{c} - \frac{ 2 \alpha } { \mu ( \alpha \sigma + 1 ) \theta } \bar{D} ( \ast \hat{t}_{c} ) \nonumber \\
& & - \frac{ 2 \alpha^{2} }{ \mu^{2} ( \alpha \sigma + 1 )^{3} \theta^{2} } \epsilon_{abc} ( \bar{Y}^{a} \wedge \ast \hat{t}^{b} - \bar{Y}^{b} \wedge \ast \hat{t}^{a} ) \nonumber \\
& & - \frac{ 2 \alpha^{3} }{ \mu^{2} ( \alpha \sigma + 1 )^{3} \theta^{2} } \epsilon_{abc} \ast \hat{t}^{a} \wedge \ast \hat{t}^{b} \nonumber \\
& & + \left( 1 + \frac{ \alpha^{2} ( i \bar{\psi} \psi ) }{ \mu ( \alpha \sigma + 1 )^{2} \theta } - \frac{ 2 \alpha^{2} \tilde{\Lambda} }{ \mu^{2} ( \alpha \sigma + 1 )^{3} \theta^{2} } \right) t_{c} \, . 
\label{eqn_2_23}
\end{eqnarray}  
Notable feature of spinor-matter coupled equation is that the coefficients $ \beta_{i} $ ($ i=1,2,3,4 $) are not constants but are real scalar functions of spinor field. This is indeed remarkable when one makes a comparison with connection-independent matter coupling where the coefficients are constants that depend on the parameters of MMG. Furthermore, looking at the source term (\ref{eqn_2_23}), it is seen that source term involves terms that are quadratic in spinor field as well as its covariant derivative and as a consequence it involves the terms quartic in Dirac field as well. In addition, there exist terms where the spinor field and its covariant derivatives are coupled to Schouten 1-form. 

Finally, we aim to obtain the consistency relation in order that spinor-matter coupled MMG equation (\ref{eqn_2_21}) be consistent. For that, we will act the covariant derivative operator (associated with torsion-free connection) on the MMG equation and use Bianchi identities as done in the previous section. Acting the covariant derivative operator and noting that the coefficients $ \beta_{i} $ ($ i = 1,2,3,4 $) are not constants but are functions of spinor fields, one obtains
\begin{eqnarray}
\epsilon_{abc} \, \bar{D} \beta_{1} \wedge \bar{R}^{ab} + \bar{D} \beta_{2} \wedge \bar{C}_{c} + \bar{D} \beta_{3} \wedge \ast e_{c} \nonumber \\
+ \frac{ 1 }{2} \epsilon_{abc} \,  \bar{D} \beta_{4} \wedge \bar{Y}^{a} \wedge \bar{Y}^{b} + \beta_{4} \epsilon_{abc} \, \bar{C}^{a} \wedge \bar{Y}^{b} + \bar{D} \tilde{t}_{c} = 0 \, .
\end{eqnarray} 
Now eliminating Cotton 2-form $ \bar{C}^{a} $ from spinor-matter coupled MMG equation (\ref{eqn_2_21}) (for on-shell consistency) and using the identities (\ref{eqn_1_32}) together with curvature-Schouten identity (\ref{eqn_1_26}), one finally gets the consistency relation in the compact form given by
\begin{eqnarray}
\bar{D} \tilde{t}_{c} = \frac{ \beta_{4} }{ \beta_{2} } \epsilon_{abc} \, \tilde{t}^{a} \wedge \bar{Y}^{b} + \frac{ 1 }{ \beta_{2} } \bar{D} \beta_{2} \wedge \tilde{t}_{c} \nonumber \\
  + \,  \epsilon_{abc} \, \left( \frac{ \beta_{1} }{ \beta_{2} } \bar{D} \beta_{2} - \bar{D} \beta_{1} \right) \wedge ( \bar{Y}^{a} \wedge e^{b} - \bar{Y}^{b} \wedge e^{a} ) \nonumber \\
+ \left( \frac{ \beta_{3} }{ \beta_{2} } \bar{D} \beta_{2} - \bar{D} \beta_{3} \right) \wedge \ast e_{c} + \frac{ 1 }{2} \epsilon_{abc} \, \left( \frac{ \beta_{4} }{ \beta_{2} } \bar{D} \beta_{2} - \bar{D} \beta_{4} \right) \wedge \bar{Y}^{a} \wedge \bar{Y}^{b} \, .
\label{spinor_consistency}  
\end{eqnarray} 
We conclude that the consistency of spinor-matter MMG equation (\ref{eqn_2_21}) requires the source term (\ref{eqn_2_23}) associated with spinor fields to satisfy the relation (\ref{spinor_consistency}). It is clearly seen that compared to consistency relation (\ref{consistency}) for the connection-independent matter coupling case, the form of the consistency equation for spinor-matter coupled MMG involves non-trivial derivative terms due to non-constant coefficients present in the MMG equation (\ref{eqn_2_21}). The relation (\ref{spinor_consistency}) can also be interpreted as the condition that should be satisfied by the spinor field in order that MMG equation be consistent. Now let us calculate the left and the right hand sides of the consistency relation (\ref{spinor_consistency}) explicitly in order to see that for the source 2-form given in (\ref{eqn_2_23}) whether both sides are identically equal (as in the connection-independent matter coupling) or an additional relation should be imposed on the spinor field in order that the minimal coupling be consistent. Before performing the calculations, first it would be convenient to express spinor field equations (\ref{eqn_2_11}) and (\ref{eqn_2_12}) in more simplified forms
\begin{equation}
\ast e^{a} \wedge \gamma_{a} \bar{D} \psi = \left( \frac{ 1 }{2} \alpha \lambda + m \right) \psi \ast 1 
\label{spinor_equation}
\end{equation}
and 
\begin{equation}
\ast e^{a} \wedge \bar{D} \bar{\psi} \gamma_{a} = - \left( \frac{ 1 }{2} \alpha \lambda + m \right) \bar{ \psi } \ast 1
\label{spinor_conjugate_equation}
\end{equation}
where using expression (\ref{eqn_2_20}) for $ \lambda^{a} $ 
\begin{equation}
\lambda = \iota_{a} \lambda^{a} =  - \frac{ 1 }{ \mu ( \alpha \sigma + 1 )^{2} \theta } \left( \frac { 1 }{2} \bar{ R } + 3 \tilde{ \Lambda } + \alpha t \right) 
\label{lambda_1}
\end{equation}  
in terms of the curvature scalar $ \bar{R} $ and the trace $ t $ of Dirac stress-energy 2-form $ t_{c} $. Note that one can interpret $ \lambda $ as spinorial self-interaction term (see also \cite{fabbri} where Dirac spinor interactions are examined in $f(R)$ gravity involving torsion). Next by using the expression (\ref{eqn_2_14}) for $ t_{c} $ and spinor field equations (\ref{spinor_equation}) and (\ref{spinor_conjugate_equation}) together with $ \lambda $ expression given above, the trace $ t $ can be obtained as
\begin{equation}
t = - \left( \frac{ \alpha }{ \mu ( \alpha \sigma + 1 )^{2} } \left( \frac{ 1 }{2} \bar{R} + 3 \tilde{\Lambda} \right) + m \theta \right) ( i \bar{\psi} \psi ) \, .
\label{trace}
\end{equation}      
If one further substitutes $ t $ in expression (\ref{lambda_1}), $ \lambda $ can be explicitly calculated as
\begin{equation}
\lambda = - \frac{ 1 }{ \mu ( \alpha \sigma + 1 )^{2} } \left( \frac{ 1 }{2} \bar{R} + 3 \tilde{ \Lambda } - \alpha \, m ( i \bar{\psi} \psi ) \right) \, .
\label{lambda_2}
\end{equation} 
Before moving any further, by employing spinor field equations, let us also express the stress-energy 2-form $ t_{c} $ in a more convenient form as
\begin{equation}
t_{c} = \rho_{c} + \frac{ 1 }{2} \alpha \lambda ( i \bar{\psi} \psi ) \ast e_{c} 
\label{dirac_stress_energy_1}
\end{equation}
where $ \rho_{c} $ explicitly reads
\begin{equation}
\rho_{c} = \frac{ i }{2} \left( \ast e^{a} \wedge \iota_{c} \bar{D} \bar{\psi} \, \gamma_{a} \, \psi - \ast e^{a} \wedge \bar{\psi} \, \gamma_{a} \, \iota_{c} \bar{D} \psi \right) \, .
\label{dirac_stress_energy_2}
\end{equation}
By using spinor field equations, one can show that $ \bar{D} \rho_{c} = 0 $. Then, if one further identifies $ \xi : = i \bar{\psi} \psi $, one can deduce that 
\begin{equation}
\bar{D} t_{c} = \frac{ 1 }{2} \alpha \bar{D} ( \lambda \, \xi ) \wedge \ast e_{c} \, .
\label{covariant_derivative}
\end{equation}  
Then for the source term (\ref{eqn_2_23}), the left hand side of the consistency equation (\ref{spinor_consistency}) can be calculated as
\begin{eqnarray}
\bar{D} \tilde{t}_{c} = - \frac{ 2 }{ \mu ( \alpha \sigma + 1 ) \theta^{2} } \bar{D} \theta \wedge \bar{C}_{c} - \frac{ 2 \alpha }{ \mu ( \alpha \sigma + 1 ) \theta } \bar{R}_{cb} \wedge \ast \hat{t}^{b} \nonumber \\
+ \frac{ 8 \alpha^{2} }{ \mu^{2} ( \alpha \sigma + 1 )^{3} \theta^{3} } \epsilon_{abc} \bar{D} \theta \wedge \bar{Y}^{a} \wedge \ast \hat{t}^{b} \nonumber \\
- \frac{ 4 \alpha^{2} }{ \mu^{2} ( \alpha \sigma + 1 )^{3} \theta^{2} } \epsilon_{abc} \left( \bar{C}^{a} \wedge \ast \hat{t}^{b} 
- \bar{Y}^{a} \wedge \bar{D} ( \ast \hat{t}^{b} ) \right) \nonumber \\
+ \frac{ 4 \alpha^{3} }{ \mu^{2} ( \alpha \sigma + 1 )^{3} \theta^{3} } \epsilon_{abc} \bar{D} \theta \wedge \ast \hat{t}^{a} \wedge \ast \hat{t}^{b} - \frac{ 4 \alpha^{3} }{ \mu^{2} ( \alpha \sigma + 1 )^{3} \theta^{2} } \epsilon_{abc} \bar{D} ( \ast \hat{t}^{a} ) \wedge \ast \hat{t}^{b} \nonumber \\
+ \frac{ \alpha^{2} }{ \mu ( \alpha \sigma + 1 )^{2} \theta^{2} } \left( \frac{ 1 }{ 2 } - \frac{ 4 \alpha^{2} \tilde{\Lambda} }{ \mu^{2} ( \alpha \sigma + 1 )^{3} \theta } \right) \bar{D} \xi \wedge t_{c} \nonumber \\
+ \frac{ 1 }{ 2 } \alpha \left( 1 + \frac{ \alpha^{2} \xi }{ \mu ( \alpha \sigma + 1 )^{2} \theta } - \frac{ 2 \alpha^{2} \tilde{ \Lambda } }{ \mu^{2} ( \alpha \sigma + 1 )^{3} \theta^{2} } \right) \bar{D} ( \lambda \xi ) \wedge \ast e_{c} 
\end{eqnarray}  
where we have used 
\begin{equation}
\bar{D}^{2} ( \ast \hat{t}_{c} ) = \bar{R}_{cb} \wedge \ast \hat{t}^{b} \, . 
\end{equation}      
Next we eliminate the Cotton 2-forms $ \bar{C}_{c} $ and $ \bar{C}^{a} $ from the spinor-matter coupled MMG field equation (\ref{eqn_2_21}) and then use the curvature-Schouten identity (\ref{eqn_1_26}) together with the identities
\begin{eqnarray}
\epsilon_{abc} t^{a} \wedge \bar{Y}^{b} &=& \bar{Y}_{b} \wedge e_{c} \wedge \ast t^{b} \nonumber \\
& = & \bar{Y}_{b} \wedge e_{c} \wedge \ast \hat{t}^{b} 
\label{spinor_stress_energy_identity_1}
\end{eqnarray}
and
\begin{eqnarray}
\bar{Y}_{c} \wedge e_{b} \wedge \ast \hat{t}^{b} & = & \bar{Y}_{c} \wedge e_{b} \wedge \ast t^{b} \nonumber \\
&=& - \iota_{b} \rho^{b} \wedge \ast \bar{Y}_{c} \nonumber \\
& = & \frac{ 1 }{ 2 } \bar{D} \xi \wedge \ast \bar{Y}_{c}
\label{spinor_stress_energy_identity_2} 
\end{eqnarray}
satisfied by Dirac stress-energy 2 form $ t_{c} $. After tedious calculations, one finally obtains the following expressions for the left and the right hand sides of the consistency relation (\ref{spinor_consistency}). Then the left hand side becomes 
\begin{eqnarray}
& & LHS = \bar{D} \tilde{t}_{c} = \frac{ 1 }{ \beta_{2} } \bar{D} \beta_{2} \wedge \left( \beta_{1} \epsilon_{abc} \bar{R}^{ab} + \beta_{3} \ast e_{c} + \frac{ 1 }{2} \beta_{4} \epsilon_{abc} \bar{Y}^{a} \wedge \bar{Y}^{b} + \tilde{t}_{c} \right) \nonumber \\
& & + \frac{ 2 \alpha }{ \mu ( \alpha \sigma + 1 )^{2} \theta^{2} } \left( 1 - \frac{ 2 \alpha^{2} \tilde{ \Lambda } }{ \mu^{2} ( \alpha \sigma + 1 )^{3} \theta } \right) \bar{Y}_{b} \wedge e_{c} \wedge \ast t^{b} \nonumber \\  
& & + \frac{ 4 \alpha^{2} }{ \mu^{2} ( \alpha \sigma + 1 )^{3} \theta^{3} } \epsilon_{abc} \bar{D} \theta \wedge \bar{Y}^{a} \wedge \ast \hat{t}^{b} - \frac{ 4 \alpha^{2} }{ \mu^{2} ( \alpha \sigma + 1 )^{3} \theta^{2} } \frac{ \beta_{4} }{ \beta_{2} } \bar{Y}_{c} \wedge \ast \hat{t}_{b} \wedge \bar{Y}^{b} \nonumber \\
& & + \frac{ 1 }{ \beta_{2} } \frac{ 16 \alpha^{4} }{ \mu^{4} ( \alpha \sigma + 1 )^{6} \theta^{4} } \ast \hat{t}_{c} \wedge \ast \hat{t}_{b} \wedge \bar{Y}^{b} - \frac{ 4 \alpha^{2} }{ \mu^{2} ( \alpha \sigma + 1 )^{3} \theta^{2} } \epsilon_{abc} \bar{D} ( \ast \hat{t}^{a} ) \wedge \bar{Y}^{b} \nonumber \\
& & + \nu_{1} \bar{D} \xi \wedge \ast \bar{P}_{c} + \nu_{2} \bar{D} \xi \wedge \ast e_{c} + \nu_{3} \bar{D} \xi \wedge \rho_{c} + \nu_{4} \bar{D} \bar{R} \wedge \ast e_{c} 
\label{left_hand_side}
\end{eqnarray}
where we have also used
\begin{equation}
\bar{D} \beta_{2} = \frac{ 2 }{ \mu ( \alpha \sigma + 1 ) \theta^{2} }{ \bar{D} \theta } 
\end{equation}
and the non-constant coefficients $ \nu_{i} $'s read 
\begin{equation}
\nu_{1} = \frac{ \alpha }{ \mu ( \alpha \sigma + 1 )^{2} \theta^{2} } \left( \frac{ 2 \alpha^{2} \tilde \Lambda }{ \mu^{2} ( \alpha \sigma  + 1 )^{3} \theta } - 1 \right) \, ,
\end{equation}
\begin{eqnarray}
\nu_{2} &=& - \frac{ \alpha^{2} \beta_{3} }{ \mu ( \alpha \sigma + 1 )^{2} \theta } + \frac{ \alpha }{ 2 \mu ( \alpha \sigma + 1 )^{2} \theta^{2} } \left( 1 - \frac{ 2 \alpha^{2} \tilde{ \Lambda } }{ \mu^{2} ( \alpha \sigma + 1 )^{3} \theta } \right) \left( \frac{ \bar{R} }{2} + \alpha m \xi \right) \nonumber \\
& & - \frac{ \alpha }{ 2 \mu ( \alpha \sigma + 1 )^{2} \theta } \left( 1 - \frac{ 2 \alpha^{2} \tilde{ \Lambda } }{ \mu^{2} ( \alpha \sigma + 1 )^{3} \theta } \right) \left( \frac{ \bar{R} }{2} + 6 \tilde{ \Lambda } - 2 \alpha m \xi - 3 \alpha \Lambda \right) \, , \nonumber \\    
\end{eqnarray}
\begin{equation}
\nu_{3} = \frac{ \alpha^{2} }{ \mu ( \alpha \sigma + 1 )^{2} \theta^{2} } \left( 1 - \frac{ 2 \alpha^{2} \tilde{ \Lambda } }{ \mu^{2} ( \alpha \sigma + 1 )^{3} \theta } \right) 
\end{equation}
and
\begin{equation}
\nu_{4} = - \frac{ \alpha \xi }{ 4 \mu ( \alpha \sigma + 1 )^{2} \theta } \left( 1 - \frac{ 2 \alpha^{2} \tilde{ \Lambda } }{ \mu^{2} ( \alpha \sigma + 1 )^{3} \theta } \right) \, . 
\end{equation} 
For the right hand side, we substitute source 2-form $ \tilde{t}^{a} $, use the identity (\ref{spinor_stress_energy_identity_1}) and make necessary simplifications to obtain the expression
\begin{eqnarray}
RHS & = & \frac{ 4 \alpha^{2} }{ \mu^{2} ( \alpha \sigma + 1 )^{3} \theta^{3} } \epsilon_{abc} \bar{D} \theta \wedge \ast \hat{t}^{a} \wedge \bar{Y}^{b} - \frac{ 4 \alpha^{2} }{ \mu^{2} ( \alpha \sigma + 1 )^{3} \theta^{2} } \epsilon_{abc} \bar{D} ( \ast \hat{t}^{a} ) \wedge \bar{Y}^{b} \nonumber \\
& & - \frac{ 4 \alpha^{2} }{ \mu^{2} ( \alpha \sigma + 1 )^{3} \theta^{2} } \frac{ \beta_{4} }{ \beta_{2} } \bar{Y}_{c} \wedge \ast \hat{t}_{b} \wedge \bar{Y}^{b} + \frac{ 1 }{ \beta_{2} } \frac{ 16 \alpha^{4} }{ \mu^{4} ( \alpha \sigma + 1 )^{6} \theta^{4} } \ast \hat{t}_{c} \wedge \ast \hat{t}_{b} \wedge \bar{Y}^{b} \nonumber \\
& & + \frac{ 2 \alpha }{ \mu ( \alpha \sigma + 1 )^{2} \theta^{2} } \left( 1 - \frac{ 2 \alpha^{2} \tilde{\Lambda} }{ \mu^{2} ( \alpha \sigma + 1 )^{3} \theta } \right) \bar{Y}_{b} \wedge e_{c} \wedge \ast t^{b} \nonumber \\
& & + \frac{ 1 }{ \beta_{2} } \bar{D} \beta_{2} \wedge \left( \beta_{1} \epsilon_{abc} \bar{R}^{ab} + \beta_{3} \ast e_{c} + \frac{ 1 }{2} \beta_{4} \epsilon_{abc} \bar{Y}^{a} \wedge \bar{Y}^{b} + \tilde{t}_{c} \right) \nonumber \\
& & + \bar{\nu}_{1} \bar{D} \xi \wedge \ast \bar{P}_{c} + \bar{\nu}_{2} \bar{D} \xi \wedge \ast e_{c}   
\label{right_hand_side}
\end{eqnarray} 
where 
\begin{equation}
\bar{\nu}_{1} = \frac{ \alpha }{ \mu ( \alpha \sigma + 1 )^{2} \theta^{2} } \left( 1 - \frac{ 2 \alpha^{2} \tilde{\Lambda} }{ \mu^{2} ( \alpha \sigma + 1 )^{3} \theta } \right) = - \nu_{1} 
\end{equation}
and
\begin{eqnarray}
\bar{\nu}_{2} & = & - \frac{ \alpha }{ 2 \mu ( \alpha \sigma + 1 )^{2} \theta^{2} } \left( 1 - \frac{ 2 \alpha^{2} \tilde{\Lambda} }{ \mu^{2} ( \alpha \sigma + 1 )^{3} \theta } \right) \bar{R} \nonumber \\
& & + \frac{ \alpha \tilde{ \Lambda } }{ 2 \mu ( \alpha \sigma + 1 )^{2} \theta^{2} } \left( \frac{ 4 \alpha^{2} \tilde{ \Lambda } }{ \mu^{2} ( \alpha \sigma + 1 )^{3} \theta } - 1 \right) - \frac{ \alpha \xi }{ 4 ( \alpha \sigma + 1 ) \theta } \, .
\end{eqnarray}
Then looking at the expressions (\ref{left_hand_side}) and (\ref{right_hand_side}) on the left and the right hand sides of the consistency relation, it can remarkably be seen that both sides become identically equal either when $ \xi = i \bar{\psi} \psi = 0 $ or when 
\begin{equation}
\xi = i \bar{\psi} \psi = \frac{ 2 \left( \mu^{2} ( \alpha \sigma + 1 )^{3} - 2 \alpha^{3} \Lambda \right) }{ 3 \alpha^{2} \mu ( \alpha \sigma + 1 ) } : = c
\label{constant_c}
\end{equation}    
where for both values of $ \xi $, it becomes $ \nu_{4} = 0 $. Also recall that for both cases, since $ \bar{D} \xi = 0 $, we have 
$ \bar{D} \beta_{i} = 0 $ and $ \bar{D} \theta = 0 $ as well.  

\noindent Notice that the case when $ \xi = 0 $ is interesting since for $ \xi = 0 $, the trace $ t $ of Dirac stress-energy 2-form 
$ t_{c} $ vanishes. In fact, it is not unusual since in literature there exist some works which study the exact solutions of Einstein-Cartan-Dirac theories in $ 3 $-dimensional and $ 4 $-dimensional spacetimes and consider the special case with $ \xi = i \bar{\psi} \psi = 0 $  (see also \cite{adak,seitz,dimakis,baekler,dereli_tucker}). As a remark, we also note that the case $ \xi = 0 $ implies that taking the representation for the spinor field in the form 
\begin{equation}
\psi = \left( 
\begin{array}{c}
\psi_{1} \\
\psi_{2} 
\end{array} \right)
\end{equation}
and using the representation (\ref{representation}) for $ \gamma_{0} $, one obtains 
\begin{equation}
i \bar{\psi} \psi = i \psi^{\dagger} \gamma_{0} \psi = | \psi_{2} |^{2} - | \psi_{1} |^{2} = 0
\end{equation}
which further implies the relation $ \psi_{2} = e^{ i \varphi } \psi_{1} $ between the components of spinor field up to a phase 
$ \varphi $. 

\noindent Then, it can be concluded that for both fixed values of $ \xi $  presented above, the left and right sides of the consistency relation are identically equal so that the minimal spinor-matter coupling is consistent either when $ \xi = 0 $ or $ \xi = c $ where non-zero constant $ c $ is given by (\ref{constant_c}). 

On the other hand, for $ \xi \neq 0 $ (and also for $ \xi \neq c $), if one compares the left and the right hand sides of the consistency relation, it is seen that the first six terms on both sides become identically equal noticing that the first term on the right hand side can also be written in the form
\begin{equation}
\epsilon_{abc} \bar{D} \theta \wedge \ast \hat{t}^{a} \wedge \bar{Y}^{b} = \epsilon_{abc} \bar{D} \theta \wedge \bar{Y}^{a} \wedge \ast \hat{t}^{b}
\end{equation}
with the appropriate change of the indices. However, looking at the terms involving the non-constant coefficients $ \nu_{i} $'s and $ \bar{\nu}_{i} $'s, it is obvious that $ \nu_{1} \neq \bar{\nu}_{1} $, $ \nu_{2} \neq \bar{\nu}_{2} $, $ \nu_{3} \neq 0 $ and $ \nu_{4} \neq 0 $ and as a result the left and the right hand sides are not identically equal for $ \xi \neq 0 $. It implies that in order that spinor-matter coupled MMG equation (\ref{eqn_2_21}) be consistent for $ \xi \neq 0 $, the consistency relation imposes an additional constraint between the spinor fields and the curvature terms (i.e Ricci 1-form $ \bar{P}_{c} $ and the curvature scalar $ \bar{R} $) in the form  
\begin{equation}
2 \nu_{1} \bar{D} \xi \wedge \bar{P}_{c} + ( \nu_{2} - \bar{\nu}_{2} ) \bar{D} \xi \wedge \ast e_{c} + \nu_{3} \bar{D} \xi \wedge \rho_{c} + \nu_{4} \bar{D} \bar{R} \wedge \ast e_{c} = 0 \, .
\label{consistency_constraint}
\end{equation}
In fact, this constraint can be interpreted as the condition that should be satisfied by spinor field in order that the minimal spinor-matter coupling be consistent. At this point, one can also comment that in order to obtain a consistent coupling without any such additional constraint on the spinor fields, the minimal coupling assumption made at the beginning of spinor-coupling can be modified and one can consider non-minimal couplings of spinor fields. However, anticipating the forms of such non-minimal couplings is not straightforward. Moreover, non-minimal couplings could produce extra difficulties especially in obtaining an algebraic solution for the auxiliary field $ \lambda^{a} $.

Before closing this section, we further remark that if one considers the case where $ \alpha = 0 $ and $ \lambda^{a} = 0 $ in the spinor-matter coupled Lagrangian and make independent variations with respect to co-frames, connections and spinor fields, one gets the field equations of spinor-matter coupled TMG theory that is also discussed in \cite{adak} (except that there also exists Mielke-Baekler term of the form $ e_{a} \wedge T^{a} $ \cite{mielke}, in the Lagrangian presented in related work).

\section{ Conclusion } 

In this work, by using the exterior algebra formalism, we have examined the matter coupling in MMG for two different cases where in the first one the matter coupling has been considered to be independent of the connection field while in the second, a general connection-dependent matter coupling has been investigated. For the case where the matter coupling is connection independent, we have obtained MMG field equation with the source 2-form that involves quadratic terms associated with stress-energy 2-form as well as its covariant derivative. We have also derived the consistency relation that the source 2-form should satisfy in order that MMG equation be consistent on shell. Also, we have presented some connection-independent matter Lagrangians to better illustrate the use of the exterior algebra notation. Next, we have considered a general connection-dependent matter coupling and obtained the field equations for this case. We have also discussed that to obtain the MMG field equation for the general matter coupling, it is required to solve the auxiliary field $ \lambda^{a} $ analytically. However, we have commented that the algebraic solution for the auxiliary field may not be possible for most of the cases depending on the form of connection-dependent matter Lagrangian. On the other hand, we have claimed that if the matter Lagrangian involves terms that are of the first order in connection then the auxiliary field can be solvable analytically. We have illustrated this for Dirac spinor-matter Lagrangian and in the rest of the work concentrated on the spinor-matter coupling.  
Then by solving the auxiliary field in terms of the stress-energy 2-form associated with spinor fields, we have derived the MMG equation while constructing the source term as well. Remarkably, we have observed that the coefficients of MMG equation are not constants but functions of spinor fields in contrast to MMG equation with connection-independent matter coupling. 
On the other hand, we have seen that similar to connection-independent matter coupling, the source term contains terms that are quadratic in spinor-matter stress-energy 2-form. There also exist covariant derivative term associated with stress-energy 2-form and additional terms that contain the covariant derivatives of spinor fields. Finally, we have derived the consistency relation  that the source term should satisfy in order that MMG equation be consistent. We have noted that the form of the consistency relation for the spinor-matter coupling considerably differs from the consistency relation for connection-independent matter coupling such that the consistency relation of spinor-matter coupled MMG contains derivative terms owing to non-constant coefficients of MMG equation.           

As closing remarks, we expect that exterior algebra formalism can enable to construct supersymmetric version of minimal massive 3d gravity (i.e minimal massive 3d supergravity). Also we note that by using this formalism, one can examine (minimal) matter couplings in other (exotic) 3d massive gravity models presented in \cite{afshar_1}, \cite{ozkan}, \cite{afshar_2} and \cite{geiller}. 
In addition, for future works, we consider to obtain some exact solutions (such as cosmological solutions , pp-waves, black holes etc.) for matter-coupled MMG theory. Specifically, one can consider cosmological solutions of spinor-matter-coupled MMG theory. In addition, $pp$-wave solutions can be investigated for Maxwell-Chern-Simons matter Lagrangians coupled to MMG where an exact self-dual solution is presented for TMG coupled to Maxwell-Chern-Simons theory \cite{dereli_3}. These would be the subjects of future research.

\appendix

\section{The solution of auxiliary 1-form $ \lambda^{a}$}

\noindent In this part of manuscript, the solution of the expression
\begin{equation}
\lambda^{a} \wedge e^{b} - \lambda^{b} \wedge e^{a} = \Lambda^{ba}
\label{eqn_a_1}
\end{equation}
for auxiliary 1-form $ \lambda^{a} $ is presented. Assume that $ \Lambda^{ba}$ is given and it can involve in general curvature terms, co-frames and some matter fields. To solve for 1-form field, we first act interior product operator $ \iota_{a} $ on (\ref{eqn_a_1}) to obtain
\begin{equation}
\lambda^{b} = \iota_{a} \Lambda^{ba} - (\iota_{a} \lambda^{a}) e^{b}
\label{eqn_a_2} 
\end{equation} 
where we have used $ \iota_{a} e^{a} = 3 $, which is the dimension of spacetime, and identity $ \iota_{b} e^{a} = \delta_{b}^{a} $. Next, by acting interior product operator $ \iota_{b}  $ on (\ref{eqn_a_2}) yields 
\begin{equation}
\iota_{a} \lambda^{a} = \frac{1}{4} \iota_{b} \, \iota_{a} \Lambda^{ba}  \,  .
\label{eqn_a_3}
\end{equation}
As a result, one obtains (by changing the indices appropriately)
\begin{equation}
\lambda^{a} = \iota_{b} \Lambda^{ab} - \frac{1}{4} ( \iota_{n} \, \iota_{p} \Lambda^{np} ) e^{a} \, . 
\label{eqn_a_4}
\end{equation}

\end{document}